\documentclass[prd,a4paper,showpacs,preprint,byrevtex]{revtex4}
\usepackage{graphicx}
\usepackage{dcolumn}
\usepackage{amsmath}
\usepackage{array}
\usepackage{bm}
\usepackage{textcomp}

\begin{document}
%\preprint{Preprint Universit\'{e} de Mons-Hainaut}

\title{Bound state equivalent potentials with the Lagrange mesh method}

\author{Fabien \surname{Buisseret}}
\thanks{FNRS Research Fellow}
\email[E-mail: ]{fabien.buisseret@umh.ac.be}
\author{Claude \surname{Semay}}
\thanks{FNRS Research Associate}
\email[E-mail: ]{claude.semay@umh.ac.be}
\affiliation{Groupe de Physique Nucl\'{e}aire Th\'{e}orique,
Universit\'{e} de Mons-Hainaut,
Acad\'{e}mie universitaire Wallonie-Bruxelles,
Place du Parc 20, BE-7000 Mons, Belgium}

\date{\today}

\begin{abstract}
The Lagrange mesh method is a very simple procedure to accurately solve
eigenvalue problems starting from a given nonrelativistic or
semirelativistic two-body Hamiltonian with local or nonlocal potential.
We show in this work that it can be applied to solve the inverse
problem, namely, to find the equivalent local potential starting from a
particular bound state wave function and the corresponding energy. In
order to check the method, we apply it to several cases which are
analytically solvable: the nonrelativistic harmonic oscillator and
Coulomb potential, the nonlocal Yamaguchi potential and the
semirelativistic harmonic oscillator. The potential is accurately
computed in each case. In particular, our procedure deals efficiently
with both nonrelativistic and semirelativistic kinematics.
\end{abstract}

% PACS
% 02.70.-c Computational techniques
% 03.65.Ge Solutions of wave equations: bound states
% 02.30.Mv Approximations and expansions
\maketitle

\section{Introduction}

The Lagrange mesh method is a very accurate and simple procedure to
compute eigenvalues and eigenfunctions of a two-body Schr\"{o}dinger
equation \cite{baye86,vinc93,baye95}. It is applicable for both local
and nonlocal interactions \cite{nonloc}, and also for a semirelativistic
kinetic operator, i.e. the spinless Salpeter
equation \cite{sem01,brau2}. In this method, the trial eigenstates are
developed in a basis of well-chosen functions, the Lagrange functions,
and the Hamiltonian matrix elements are obtained with a Gauss
quadrature. Moreover, the Lagrange mesh method can be extended to treat
very accurately
three-body problems, in nuclear or atomic physics \cite{hess99,3b2}.

In this work, we apply the Lagrange mesh method to solve the
inverse problem for bound states: starting from a given bound state --
wave function and corresponding eigenenergy --, we show how to compute
the
equivalent local potential. To our knowledge, this application of
Lagrange mesh method has not been studied before. It can then be used to
compute the equivalent local potential of a given nonlocal potential.
The determination of equivalent local potentials is of particular
interest in nuclear physics (see for example Ref.~\cite{nucl}). The more
interesting point is that our procedure allows to deal with
semirelativistic kinematics.

Our paper is organized as follows. In Sec.~\ref{lagmesh}, we recall
the main points of the Lagrange mesh method and show how to apply it to
solve a bound state problem with a central potential. Then, we give a
procedure to compute the equivalent local potential with this method
starting from a given spectrum in Sec.~\ref{bsep}. In order to check the
efficiency of our method, we apply it to several cases in which the
spectrum is analytically known. Firstly, we consider three central
potentials with a nonrelativistic kinematics in Sec.~\ref{applic}: the
harmonic oscillator (Sec.~\ref{nrho}), the Coulomb potential
(Sec.~\ref{nrcp}), and the nonlocal Yamaguchi potential
(Sec.~\ref{yama}).
Secondly, in Sec.~\ref{srho}, we consider the case of the
semirelativistic harmonic oscillator for two massless particles, whose
solution is also analytical. The accuracy of the method is checked in
all those cases, and conclusions are drawn in Sec.~\ref{conclu}.

\section{Lagrange mesh method}\label{lagmesh}

A Lagrange mesh is formed of $N$ mesh points $x_{i}$ associated with an
orthonormal set of indefinitely derivable functions $f_{j}(x)$ on an
interval $[a,b]$. A Lagrange function $f_{j}(x)$
vanishes
at all mesh points but one; it satisfies the
condition \cite{baye86,vinc93,baye95}
\begin{equation}
\label{flagpro}
f_{j}(x_{i})=\lambda^{-1/2}_{i}\delta_{ij}.
\end{equation}
The weights $\lambda_{i}$ are linked to the mesh points $x_{i}$ through
a Gauss quadrature formula
\begin{equation}
\label{gauss}
\int^{b}_{a} g(x)\, dx  \approx\sum^{N}_{k=1}\lambda_{k}\, g(x_{k}),
\end{equation}
which is used to compute all the integrals over the interval $[a,b]$.

As in this work we only study radial equations, we consider the interval
$[0,\infty[$, leading
to a Gauss-Laguerre quadrature. The Gauss formula~(\ref{gauss}) is
exact when $g(x)$ is a polynomial of degree $2N-1$ at most, multiplied
by $\exp(-x)$. The $N$ Lagrange-Laguerre mesh points $x_i$ are then
given by the zeros of
the Laguerre polynomial $L_{N}(x)$ of degree $N$ \cite{baye86}.
An explicit form can be derived for the corresponding regularized
Lagrange functions
\begin{equation}
\label{flag}
f_{i}(x)=(-1)^{i}x^{-1/2}_{i}\, x(x-x_{i})^{-1}L_{N}(x)\, e^{-x/2}.
\end{equation}
They clearly satisfy the constraint ($\ref{flagpro}$), and they are
orthonormal, provided the scalar products are computed with the
quadrature  ($\ref{gauss}$). Moreover, they vanish in $x=0$.

To show how these elements can be applied to a physical problem,
let us consider a standard Hamiltonian $H = T(\vec{p}^{\, 2})+V(r)$,
where
$T(\vec{p}^{\, 2})$ is the kinetic term and $V(r)$ a radial potential
(we work in natural units $\hbar = c=1$).
The calculations are performed with trial states $|\psi\rangle$ given by
\begin{equation}
\label{state}
\left|\psi\right\rangle=\sum^{N}_{k=1}C_{k}\left|f_{k}\right\rangle,
\end{equation}
where
\begin{equation}
\left\langle \vec{r}\,|f_{k}\right\rangle=
\frac{f_{k}(r/h)}{\sqrt{h}\,r}Y_{\ell m}(\theta,\varphi).
%\equiv R(r)Y_{\ell m}(\hat{r}).
\end{equation}
$\ell$ is the orbital angular momentum quantum number and the
coefficients $C_{k}$ are
linear variational parameters. $h$ is a scale parameter chosen to
adjust the size of the mesh to the domain of physical interest. If we
define $r=h\,x$,
with $x$ a dimensionless variable, a relevant value of $h$ will be
obtained thanks to the relation $h=r_a/x_N$, where $x_N$ is the last
mesh point and $r_a$ is a physical radius for which the asymptotic tail
of the wave function is well defined. This radius has to be a priori
estimated, but various computations show that it has not to be known
with great accuracy, since the method is not variational in
$h$ \cite{sem01,fab1}.

We have now to compute the Hamiltonian matrix elements. Let us
begin with the potential term. Using the
properties of the Lagrange functions and the Gauss
quadrature~(\ref{gauss}), the potential matrix for a local potential
$V(r)$ is diagonal.
Its elements are
\begin{equation}
\label{poten1}
V_{ij}=\int^\infty_0 dx\,f_i(x) V(h\, x) f_j(x)\approx
V(hx_{i})\, \delta_{ij},
\end{equation}
and only involve the value of the potential at the mesh points.
As the matrix elements are computed only approximately, the variational
character of the method cannot be guaranteed. But the accuracy of the
method is preserved \cite{baye02}. The matrix elements for a nonlocal
potential $W(r,r')$ are given by \cite{nonloc}
\begin{equation}\label{poten2}
W_{ij}=h\,\int^\infty_0 dx\int^\infty_0 dx' f_i(x)\,W(hx,hx')\,f_j(x')
\approx h\, \sqrt{\lambda_i\lambda_j}\ W(hx_i,hx_j).
\end{equation}

The kinetic energy operator is generally only a function of
$\vec{p}^{\, 2}$. It is shown in Ref.~\cite{baye95} that, using the
Gauss quadrature and
the properties of the Lagrange functions, one obtains the corresponding
matrix
\begin{equation}
(\vec{p}^{\, 2})_{ij}=\frac{1}{h^{2}}\left[p^{\, 2}_{r\,
ij}+\frac{\ell(\ell+1)}{x^{2}_{i}}\delta_{ij}\right],
\end{equation}
where
\begin{equation}\label{pij_def}
p^{2}_{r\, ij}=\left\{
\begin{array}{lll}
&(-1)^{i-j}(x_{i}x_{j})^{-1/2}(x_{i}+x_{j})(x_{i}-x_{j})^{-2} &(i\neq
j),\\
&(12\,x^{2}_{i})^{-1}[4+(4N+2)\,x_{i}-x^{2}_{i}]&(i=j).
\end{array} \right.
\end{equation}
Now, the kinetic energy matrix $T(\vec{p}^{\, 2})$ can be computed with
the
following method \cite{sem01}:
\begin{enumerate}
\item Diagonalization of the matrix $\vec{p}^{\, 2}$. If $D^{2}$ is the
corresponding diagonal matrix, we have thus
$\vec{p}^{\, 2}=SD^{2}S^{-1}$, where $S$ is the transformation matrix.
\item Computation of $T(D^{2})$ by taking the function $T$ of all
diagonal elements of $D^{2}$.
\item Determination of the matrix elements
$T_{ij}$ in the
Lagrange basis by using the transformation matrix $S$:
$T(\vec{p}^{\, 2})=S\,T(D^{2})\, S^{-1}$.
\end{enumerate}
Note
that such a calculation is not exact because the number of Lagrange
functions is finite. However, it has already given good results in the
semirelativistic case, when
$T(\vec{p}^{\, 2})=\sqrt{\vec{p}^{\, 2}+m^{2}}$ \cite{sem01} or even
when $T(\vec{p}^{\, 2},r)=\sqrt{\vec{p}^{\, 2}+U^{2}(r)}$ \cite{brau2}.

The eigenvalue equation
$H\left|\psi\right\rangle=E\left|\psi\right\rangle$ reduces then to a
system of $N$ mesh equations,
\begin{equation}\label{meq}
\sum^{N}_{j=1}\left[ T_{ij}+{\cal V}_{ij}-E \delta_{ij}\right]
C_{j}=0 \quad \text{with} \quad C_{j}=\sqrt{h\lambda_{j}}\, u(hx_{j}),
\end{equation}
where $u(r)$ is the regularized radial wave function and ${\cal V}$ the
local or nonlocal potential matrix. The coefficients
$C_{j}$ provide the values of the radial wave function at mesh points.
But contrary to some other mesh methods, the wave function is also known
everywhere thanks to Eq.~(\ref{state}).

\section{Bound state equivalent local potential}\label{bsep}

In the previous section, we applied the Lagrange mesh method to solve
the eigenequation for two-body central problems. We now show that this
method allows to solve very easily the inverse problem, that is,
starting from particular wave function $\left|\psi\right\rangle$ and
energy $E$, to find the corresponding equivalent local potential for a
given kinematics $T$.

In the case of a local central potential, the mesh
equations~(\ref{meq}) can be rewritten as
\begin{equation}\label{effpot1}
  V(hx_i)=E-\frac{1}{\sqrt{\lambda_i}\, u(hx_i)}\sum^N_{j=1}T_{ij}\sqrt{
  \lambda_j}\ u(hx_j).
\end{equation}
We see from the above equation that, provided we know the radial wave
function and the energy of the state, the equivalent local potential can
be directly computed at the mesh points. Let us note that, since the
matrix elements $T_{ij}$ depend on the orbital angular momentum $\ell$,
this quantum number has to be a priori specified. The calculation is
done easily because the potential matrix for a local potential $V(r)$ is
diagonal and only involves the value of the potential at the mesh
points, as shown in Eq.~(\ref{poten1}). Obviously, this method does not
require a given normalization for the wave function. Moreover, it is
also applicable for semirelativistic kinematics.

We can remark that Eq.~(\ref{effpot1}) contains term which are
proportional to $u(hx_j)/u(hx_i)$. They may be difficult to compute
numerically with a great accuracy when $h x_i$ is either close to zero
or very large. In these cases indeed, the regularized wave function
tends towards zero. It means that the first values of the potential and
also the last ones could be inaccurate. It is worth mentioning that, for
radially excited states, a particular mesh point $x_k$ could be such
that $hx_k$ is a zero of the wave function. In this case, $V(hx_k)$
cannot be computed. Although very improbable, this problem could simply
be cured by taking a slightly different value of $N$ or $h$.

In order to check the validity of our method, we will consider four
cases where the eigenvalue problem is analytically solvable for a given
potential $V^E$. This will enable us to compare the numerically computed
points $V(hx_i)$ with the corresponding exact values $V^E(hx_i)$. The
number $\delta$, defined by
\begin{eqnarray}\label{ddef}
  \delta=\max\left\{\left|\frac{V(hx_i)-V^E(hx_i)}{V^E(hx_i)}\right|,\
  3\leq i\leq N-3 \right\},
\end{eqnarray}
is a measurement of the accuracy of the numerical computations. The more
$\delta$ is close to zero, the more the method is accurate. The first
and last two mesh points are -- arbitrarily -- not included in the
computation of $\delta$, since they can introduce errors which are not
due to the method itself, but rather to a lack of precision in the
numerical computations, as we argued previously from inspection of
formula~(\ref{effpot1}).

\section{Nonrelativistic Applications}\label{applic}

The kinetic operator which will be used in all the computations of this
section is given by
\begin{equation}
T(\vec p^{\,2})=\frac{\vec p^{\,2}}{2\mu},
\end{equation}
where $\mu$ is the reduced mass of the studied two-body system.

\subsection{Harmonic oscillator}\label{nrho}

The spectrum of a spherical harmonic oscillator, whose potential reads
\begin{equation}\label{ehp}
V^E(r)=\frac{\Lambda^2r^2}{2\mu},
\end{equation}
 is given by (see for example Ref.~\cite[problem 66]{flu})
\begin{equation}\label{nrho1}
  R_{n\ell}(r)\propto r^\ell\, {\rm e}^{-\Lambda r^2/2} L^{\ell+1/2}_n(
  \Lambda\, r^2), \quad  E_{n\ell}=\Lambda\,\mu^{-1}(2n+\ell+3/2).
\end{equation}

It is readily computed from the virial theorem that
$\left\langle r^2\right\rangle=(2n+\ell+3/2)/\Lambda$. Therefore, we
suggest the following value for the scale parameter:
\begin{eqnarray}
  h&=&\frac{4\sqrt{\left\langle r^2\right\rangle}}{x_N}\label{hdef1}\\
  &=&\frac{4}{x_N}\sqrt{\frac{(2n+\ell+3/2)}{\Lambda}}\label{hdef1b},
\end{eqnarray}
where the factor $4$ ensures that the last mesh point will be located in
the asymptotic tail of the wave function.

\begin{figure}[t]
\includegraphics*[width=8.7cm]{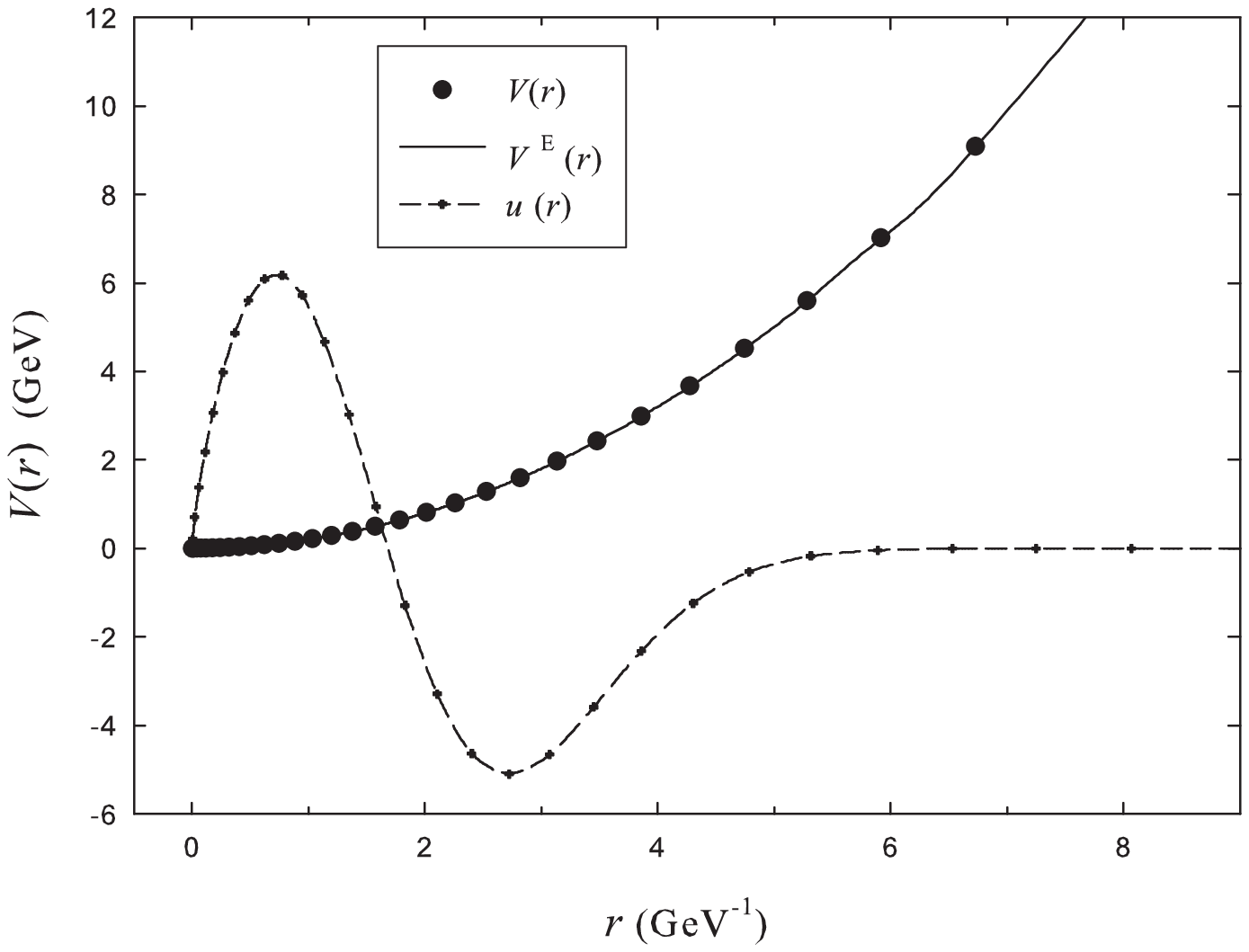}
\caption{Comparison between the potential computed from the $2S$ wave
function~(\ref{nrho1}) thanks to the Lagrange mesh
formula~(\ref{effpot1}), and the exact harmonic potential given by
Eq.~(\ref{ehp}). The equivalent potential is only known at the mesh
points (circles), and the exact potential is plotted with a solid line.
The regularized wave function is also plotted with an arbitrary
normalization (dashed line). We used $\mu=0.70$~GeV, $\Lambda=0.53$~GeV
$^2$, $N=30$, and $h=9.8~10^{-2}$~GeV$^{-1}$ following
formula~(\ref{hdef1b}).}
\label{Fig1}
\end{figure}

In order to make explicit computations, we have to specify the
value of our parameters. We set $\mu=0.70$~GeV and
$\Lambda=0.53$~GeV$^2$. These parameters can be used in hadron physics
to roughly describe
a $c\bar c$ meson \cite{fab1,drg}. We choose $N=30$, and the scale
parameter is computed by using Eq.~(\ref{hdef1b}). Once these parameters
are fixed, Eqs.~(\ref{effpot1}) and (\ref{nrho1}) allow to find the
equivalent local potential. The result is plotted and compared to the
exact harmonic potential~(\ref{ehp}) in Fig.~\ref{Fig1}, where we used
the wave function in the $2S$ state ($n=1,\,\ell=0$). The numerical
result is clearly close to the exact result, and only $30$ mesh points
are enough to provide a good picture of the potential: we have indeed
$\delta=2.1~10^{-3}~\%$, this number being computed with
Eq.~(\ref{ddef}). The same conclusion holds if other states than the
$2S$ one are used, and $\delta$ is always smaller than $1~\%$.

\begin{figure}[h]
\includegraphics*[width=9cm]{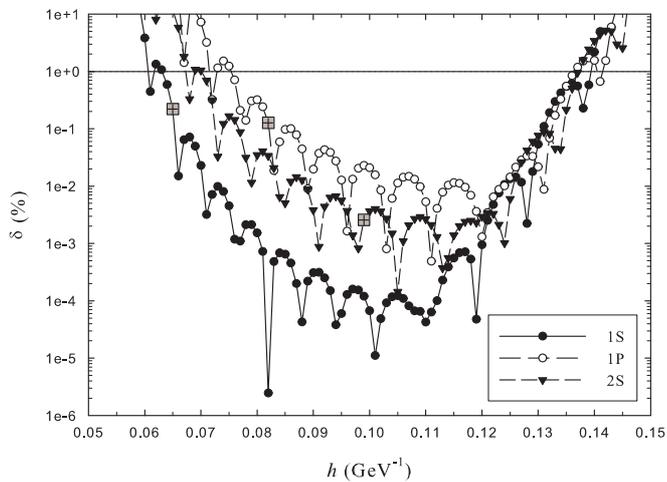}
\caption{Evolution of $\delta$ versus the scale parameter $h$ for the
$1S$ (full circles), $1P$ (empty circles), and $2S$ (triangles) states
for $N=30$. The gray boxes are the different values of $\delta$ for a
scale parameter computed with formula~(\ref{hdef1b}). They all ensure a
value of~$\delta$ lower than $1~\%$. }
\label{Fig2}
\end{figure}

In Fig.~\ref{Fig2}, we show the variation of $\delta$ with the
scale parameter $h$ for three different states and $N=30$. We can
conclude from this figure that a rather large interval exists where the
quantity $\delta$ is lower than $1~\%$. Consequently, the scale
parameter
does not need to be computed with great accuracy: our
criterion~(\ref{hdef1}) is clearly accurate enough since the predicted
value of $h$ is
always located in this interval. The global behavior of $\delta$ which
can be observed in Fig.~\ref{Fig2} is due the difficulty of computing
$V(hx_i)$ when the scale parameter is too small or too large. In this
case indeed, the mesh points $hx_i$ cover no longer the main part of the
wave function, and a partial knowledge of the wave function leads to an
inaccurate description of the potential.

\subsection{Coulomb potential}\label{nrcp}

This case is of interest since it enables us to check whether the method
we present can correctly reproduce a singular potential or not. The
radial wave function and eigenenergies of a central Coulomb potential
\begin{equation}\label{coul}
V^{E}=-\frac{\kappa}{r}
\end{equation}
respectively read (see for example Ref.~\cite[problem 67]{flu})
\begin{equation}\label{nrc}
  R(r)\propto r^\ell\,{\rm e}^{-\gamma r}\, L^{2\ell+1}_{n_p-\ell-1}(2
  \gamma r),\quad E_{n_p}=-\frac{\mu\kappa^2}{2n^2_p},
\end{equation}
with $n_p\geq 1$, $0\leq\ell\leq n_p-1$, and $\gamma=\mu\kappa/n_p$. The
principal quantum number $n_p$ is defined by $n_p=n+\ell+1$.

It can be computed that \cite[p. 147]{landau}
\begin{equation}
\left\langle
r^2\right\rangle=\frac{n^2_p}{2\mu^2\kappa^2}\left[5n^2_p+1-3\ell(\ell+1
)\right].
\end{equation}
As the evaluation of the scale parameter given by Eq.~(\ref{hdef1})
yields good results in the harmonic oscillator case, it can be adapted
to the Coulomb potential, and $h$ is now defined as
\begin{equation}\label{hdef2}
  h=\frac{15}{x_N}\frac{n_p}{\sqrt{2}\, \mu\kappa}\sqrt{5n^2_p+1-3\ell(
  \ell+1)}.
\end{equation}
A factor $15$ is now needed because the Coulomb potential is a long-
ranged one. The wave function has thus to be known on a larger domain
than for the harmonic oscillator, since the latter potential is a
confining one.

\begin{figure}[h]
\includegraphics*[width=8.8cm]{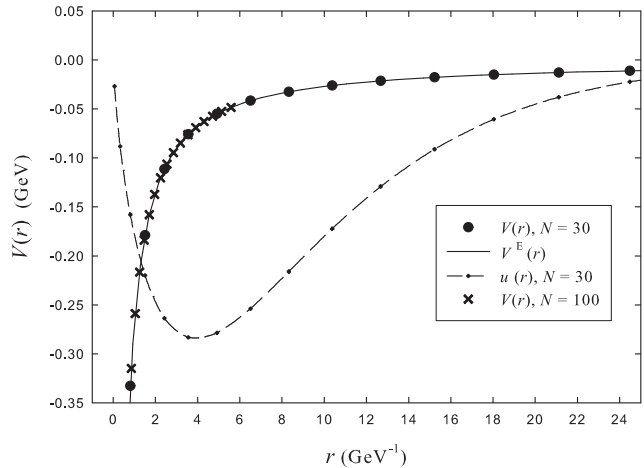}
\caption{Comparison between the potential computed from the $1S$ wave
function~(\ref{nrc}) thanks to the Lagrange mesh
formula~(\ref{effpot1}), and the exact Coulomb potential given by
Eq.~(\ref{coul}). The potential has been computed with $N=30$ (circles)
and $N=100$ (crosses) mesh points, but only a few points are plotted for
clarity. The regularized wave function is also plotted with an arbitrary
normalization (dashed line). We used $\mu=0.70$~GeV and $\kappa=0.27$.
For $N=30$, it can be computed that $h=1.3$~GeV$^{-1}$ and
$\delta=1.4~10^{-5}~\%$; for $N=100$, we have $h=0.37$~GeV$^{-1}$ and
$\delta=0.7~\%$. }
\label{Fig3}
\end{figure}

In order to numerically compute the equivalent potential from the
wave function~(\ref{nrc}), we set $\mu=0.70$~GeV and $\kappa=0.27$. The
particular value of $\kappa$ we chose is commonly used in hadron physics
to parameterize the one-gluon-exchange part of the potential between two
heavy quarks \cite{fab1}. We choose $N=30$, and the scale parameter is
computed by using Eq.~(\ref{hdef2}). The result is plotted and compared
to the exact Coulomb potential~(\ref{coul}) in Fig.~\ref{Fig3} for the
wave function in the ground state ($n=\ell=0$). The numerical result is
close to the exact result, with a value of $\delta$ which is equal to
$1.4~10^{-5}~\%$. In particular, the singular behavior is well
reproduced. To stress this point, we performed another calculation with
$N=100$, and $h=0.37$~GeV$^{-1}$ following Eq.~(\ref{hdef2}). It can be
seen in Fig.~\ref{Fig3} that the Coulomb potential is then very well
matched at short distances. In this case however, we have
$\delta=0.7~\%$. Although this precision is still very satisfactory, it
seems strange at first sight that $\delta$ is higher for a larger number
of mesh points. This is due to the fact that the mesh points are the
zeros of the Laguerre polynomial of degree $N$. The first physical point
which is taken into account in the definition of $\delta$ is $hx_3$,
which is smaller for $N=100$ ($hx_3=0.811$~GeV$^{-1}$) than for $N=30$
($hx_3=0.068$~GeV$^{-1}$). This causes $\delta$ to be larger, since the
more a point is close to zero, the more the accuracy decreases.

For what concerns the variation of $\delta$ versus $h$, the same
qualitative features than for the harmonic oscillator are observed.
Equation~(\ref{hdef2}) thus appears to give a good evaluation of the
scale parameter. It can be also checked that a factor smaller than $15$
in Eq.~(\ref{hdef2}) can lead to values of the scale parameter for which
$\delta$ is quite larger than $1~\%$.

\subsection{Yamaguchi potential}\label{yama}

The Yamaguchi potential is a separable nonlocal potential, given by
\begin{equation}\label{sepa}
  W(r,r')=-v(r)\,v(r'),
\end{equation}
with
\begin{equation}\label{vdef}
  v(r)=\sqrt{\beta/\mu}\ (\alpha+\beta)\,{\rm e}^{-\beta r}.
\end{equation}
It was introduced in Ref.~\cite{yama} to study the deuteron
($\mu=0.468$~GeV). In particular, for $\alpha=0.046$ GeV and
$\beta=0.274$~GeV, it
admits a bound state whose binding energy is the one of the deuteron,
that is $E=-2.225$~MeV.

A nice particularity of this nonlocal potential is that the bound
state wave function can be analytically determined. It reads
\begin{equation}\label{udef}
  R(r)\propto \frac{{\rm e}^{-\alpha r}-{\rm e}^{-\beta r}}{r}.
\end{equation}
Inserting this wave function into Eq.~(\ref{effpot1}) will provide us
with the equivalent local potential associated with the Yamaguchi
potential. Finding equivalent local potentials coming from nonlocal
potentials is of interest in nuclear physics, although most studies are
devoted to scattering states (see for example Refs.~\cite{nucl}). The
bound state equivalent potential of a separable nonlocal potential of
the form~(\ref{sepa}) is shown in Ref.~\cite{dijk} to be given by
\begin{equation}
  V^{L}(r)=-\frac{v(r)}{u(r)}\int^\infty_0dr'\, v(r')\,u(r'),
\end{equation}
with $u(r)$ the regularized wave function of the bound state for the
nonlocal potential. Relations~(\ref{vdef}) and (\ref{udef}) can be
injected in this last equation to compute that
\begin{equation}\label{loceq}
  V^{L}(r)=-\, \frac{\beta^2-\alpha^2}{2\mu}\, \frac{{\rm e}^{-\beta
  r}}{{\rm e}^{-\alpha r}-{\rm e}^{-\beta r}}.
\end{equation}

As the radial wave function~(\ref{udef}) is maximal in $r=0$,
$R(0)\propto(\beta-\alpha)$, we can compute the scale parameter by
demanding that
\begin{equation}
  R(hx_N)/R(0)=\epsilon,
\end{equation}
 with $\epsilon$ a small number, that we will set equal to $10^{-3}$.
 Then, assuming that $\alpha\ll\beta$ as it is the case for the
 deuteron, $h$ will approximately be given by
\begin{equation}\label{hdef3}
  h\approx-\frac{\ln\left[\epsilon(\beta-\alpha)\right]}{\alpha\, x_N}.
\end{equation}

\begin{figure}[h]
\includegraphics*[width=8.0cm]{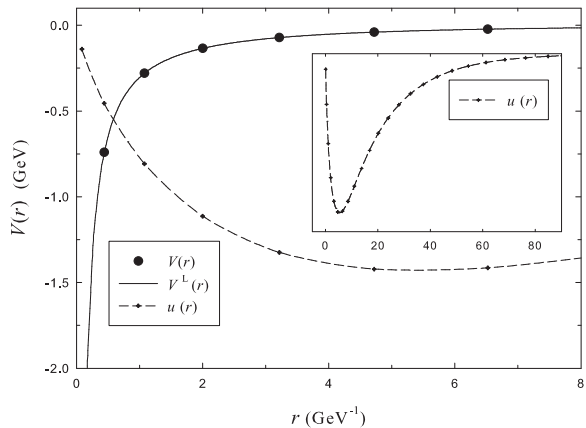}
\caption{Comparison between the equivalent local potential (circles)
computed from the wave function~(\ref{udef}) with $E=-2.225$~MeV and the
exact equivalent local potential (solid line) given by
Eq.~(\ref{loceq}). The regularized wave function is also plotted with an
arbitrary normalization (dashed line). We used $\alpha=0.046$~GeV,
$\beta=0.274$~GeV, and $N=30$. Following formula~(\ref{hdef3}),
$h=1.8$~GeV$^{-1}$.}
\label{Fig4}
\end{figure}

The equivalent local potential $V^L(r)$ and the one computed with
the Lagrange mesh method can be compared in Fig.~\ref{Fig4}. The
deuteron parameters are used, together with $N=30$ and $h$ given by
Eq.~(\ref{hdef3}). The agreement is satisfactory since
$\delta=0.31~\%$.
The extension of the wave function is large because the deuteron is
weakly bound. An estimation of its radius is indeed given by $1.96$~fm
in Ref.~\cite{bad}, that is the rather large value of $9.9$~GeV$^{-1}$.

\section{The semirelativistic harmonic oscillator}\label{srho}

A nice feature of the Lagrange mesh method is that it allows to solve
semirelativistic Hamiltonians like the spinless Salpeter equation or the
relativistic flux tube model \cite{fab1,rft}, which are relevant in
quark physics. Equation~(\ref{effpot1}) is consequently applicable if
the kinetic operator is given by
\begin{equation}
  T(\vec p^{\,2})=2\sqrt{\vec p^{\,2}+m^2}.
\end{equation}

In the ultrarelativistic case where $m=0$, the spectrum of the
Hamiltonian
\begin{equation}\label{ssh}
  H=2\sqrt{\vec p^{\,2}}+\Omega\, r^2
\end{equation}
can be analytically computed in momentum space in terms of the regular
Airy function for $\ell=0$. In position space, it reads \cite{lucha}
\begin{equation}\label{srhowf}
R(r)\propto \frac{1}{r}\int^\infty_0dp\ \sin(p\,r)\ {\rm Ai}\left[\left(
\frac{2}{\Omega}\right)^{1/3}p+\alpha_n\right],\quad
E_n=-(4\Omega)^{1/3}\alpha_n,
\end{equation}
where $\alpha_n<0$ are the zeros of ${\rm Ai}$. They can be found for
example in Ref.~\cite[table 10.13]{Abra}.

Thanks to the particular properties of the Airy function, it can be
computed that \cite{sema04}
\begin{equation}
  \left\langle r^2\right\rangle=-\left(\frac{2}{\Omega}\right)^{2/3}
  \frac{\alpha_n}{3}.
\end{equation}
The scale parameter will thus be computed with the relation
\begin{equation}\label{hdef4}
  h=\frac{4}{x_N}\left(\frac{2}{\Omega}\right)^{1/3}\sqrt{-\frac{
  \alpha_n}{3}},
\end{equation}
in analogy with the similar case of the nonrelativistic harmonic
oscillator.

\begin{figure}[h]
\includegraphics*[width=9cm]{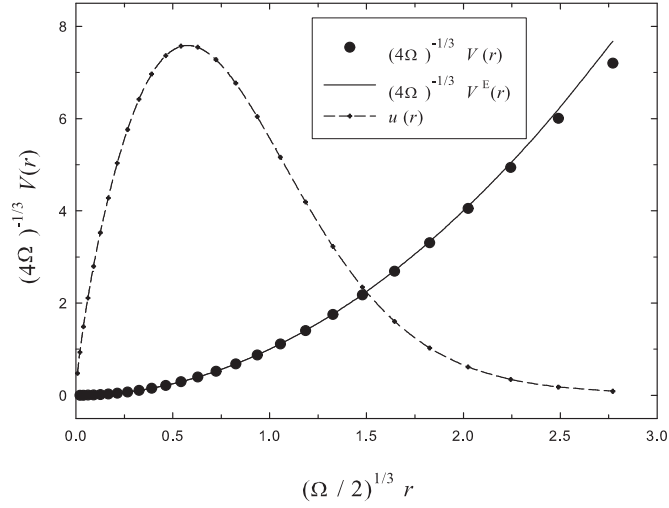}
\caption{Comparison between the potential computed from the $1S$ wave
function~(\ref{srhowf}) with a semirelativistic kinematics (circles) and
the exact harmonic potential (solid line) given by Eq.~(\ref{eho}). The
regularized wave function is also plotted with an arbitrary
normalization (dashed line). We used $N=30$, and
$(\Omega/2)^{1/3}h=0.034$ from Eq.~(\ref{hdef4}).}
\label{Fig5}
\end{figure}

The comparison between the potential computed with our method and
the exact one
\begin{equation}\label{eho}
  V^E(r)=\Omega\, r^2
\end{equation}
is given in Fig.~\ref{Fig5}. The value $\Omega=0.2$~GeV$^3$ is typical
for potential models of light quarks \cite{drg}. But, we present our
results as dimensionless quantities. The curves are thus universal: they
do not dependent on $\Omega$, which is the only parameter of this
Hamiltonian. Although still satisfactory, the agreement is not as good
as with the nonrelativistic applications. We find indeed
$\delta=3.1~\%$. By inspection of Fig.~\ref{Fig5}, it can be seen that
the last points slightly differ from the exact curve. These points are
related to the value of the wave function in its asymptotic tail, as it
can be seen from Eq.~(\ref{effpot1}). It means that finding the
equivalent potential, especially with a semirelativistic kinematics,
needs a good knowledge of the tail, which is not often necessary for
computation of the energy spectra.

\begin{figure}[h]
\includegraphics*[width=9cm]{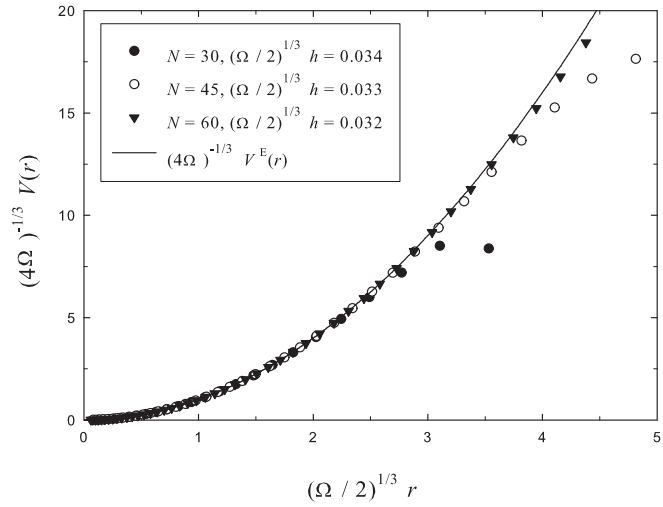}
\caption{Complete curves giving the potential computed from the $1S$
wave function~(\ref{srhowf}) with a semirelativistic kinematics for
several choices of $h$ and $N$. These choices ensure more or less the
same mesh point density. The potential is only known at the mesh points
(symbols), and the exact harmonic potential is plotted with a solid
line.}
\label{Fig6}
\end{figure}

In our case, the discrepancies for the last points are due to the
computation of the wave function in the asymptotic regime. It can be
checked that a resolution of Hamiltonian~(\ref{ssh}) with the Lagrange
mesh method leads to a wave function which asymptotically decreases
faster than the exact wave function, given by Eq.~(\ref{srhowf}).
Conversely, if one starts from the exact wave function, the Lagrange
mesh procedure will lead to a potential which does not increase enough
asymptotically, as we observe in Fig.~\ref{Fig5}. Fortunately, only the
very last points are affected, as it is shown in Fig.~\ref{Fig6}. By
varying $N$ and $h$, that is to say by varying the interval where the
potential is computed, one can always correctly reproduce the potential
in a given region: the more $hx_N$ is large, the larger is the interval
where the potential is correctly reproduced. Finding the equivalent
potential with a spinless Salpeter equation seems thus to require a more
careful study: several curves have to be computed by varying $h$ and $N$
in order to understand whether the long range behavior of the potential
is physical or simply due to a numerical artifact.

\section{Conclusions and outlook}\label{conclu}

In this work, we extended the domain of application of the Lagrange mesh
method to a particular type of problem: to find the equivalent local
potential corresponding to a given bound state with a given kinematics.
We assumed a central problem. Starting from a particular radial wave
function and the corresponding energy, the method we presented here
allows to compute the equivalent local potential at the mesh points. We
checked the accuracy of the computations in various cases whose
solutions are analytically known. Firstly, we studied the well-known
nonrelativistic harmonic oscillator and Coulomb potentials. These
potentials are correctly reproduced by the Lagrange mesh method with a
precision better than $1~\%$, provided the scale parameter is large
enough to take into account the asymptotic tail of the wave function.
Moreover, the singularity of the Coulomb potential is well matched. The
numerical parameters are the number of mesh points, and the scale
parameter. It appears that a typical value of $30$ mesh points is enough
to provide a good picture of the potential. As it was the case for usual
eigenvalue problems, the scale parameter does not need to be accurately
determined: a rather large interval exists where the precision is lower
than $1~\%$.

If the spectrum comes from a nonlocal potential, our method will
compute the equivalent local potential. This problem is of interest in
nuclear physics \cite{nucl}. As an illustration, we applied it to the
nonlocal Yamaguchi potential describing the deuteron. In this particular
case, the spectrum is analytical as well as the corresponding equivalent
potential. Again, the accuracy of our method is very good.

Finally, our procedure can also be easily adapted to the case of a
semirelativistic kinematics. As a check, we studied the semirelativistic
harmonic oscillator. Again, the potential is correctly reproduced, but
it appears that the asymptotic behavior of the potential is problematic.
This is an artifact of the method in the semirelativistic case: by
varying the mesh size, one can indeed see that the value of the
potential at the last mesh points is systematically too low, but the
harmonic shape of the potential is well reproduced at the other mesh
points.

Our purpose is to apply this method to the study of systems
containing quarks and gluons. In particular, glueballs, which are bound
states of gluons, are very interesting systems because their existence
is directly related to the nonabelian nature of QCD. Bound states of two
gluons can be described within the framework of potential models by a
spinless Salpeter equation with a Cornell potential: a linear confining
term plus a Coulomb term coming from short-range interactions
\cite{brau}. Such a phenomenological potential has been shown to arise
from QCD in the case of a quark-antiquark bound state \cite{loop}.
Theoretical indications show that it could be valid also for
glueballs \cite{simo}. Moreover, recently, the mass and the wave
function of the
scalar glueball (with quantum numbers $J^{PC}=0^{++}$) has been computed
in lattice QCD \cite{lat}. Thanks to the Lagrange mesh method, these
data could be used to extract the potential between two gluons from
lattice QCD, and see whether it is a Cornell one or not. This study will
be published elsewhere.

\acknowledgments
The authors thank the FNRS for financial support.

\newpage

\end{document}